\documentclass[oneside]{article}
\usepackage[T1]{fontenc}
\usepackage{authblk}
\usepackage{float}
\usepackage{amsmath}
\usepackage{graphicx}
\usepackage{xfrac}
\usepackage[english]{babel}
\usepackage{lmodern}
\usepackage[T1]{fontenc}
\usepackage[latin9]{inputenc}

\usepackage{csquotes}

\usepackage{mathtools}
\usepackage{graphicx}
\usepackage{esint}
\usepackage[unicode=true,
 bookmarks=false,
breaklinks=false,pdfborder={0 0 1},backref=false,colorlinks=false]
 {hyperref}
\usepackage[a4paper, total={210mm,297mm},margin=2.0cm]{geometry}

\title{Dynamic symmetry-breaking in mutually annihilating fluids with selective interfaces}

\author[1,2,3]{Sauro Succi \thanks{Electronic address: \texttt{sauro.succi@iit.it,succi@seas.harvard.edu}; Corresponding author}}

\author[2]{Andrea Montessori}

\author[4,3]{Giacomo Falcucci}

\affil[1]{Center for Life Nano Science@La Sapienza, Istituto Italiano di Tecnologia, 00161 Roma, Italy}
\affil[2]{Istituto per le Applicazioni del Calcolo CNR, via dei Taurini 19, Rome, Italy}
\affil[3]{Institute for Applied Computational Science, Harvard John A. Paulson School of Engineering And Applied Sciences, Cambridge, MA 02138, United States}
\affil[4]{Department of Enterprise Engineering ``Mario Lucertin'',University of Rome Tor Vergata, Via del Politecnico 1, 00133 Rome, Italy}

\date{\today}

\begin{document}

\maketitle

\begin{abstract}
The selective entrapment of mutually annihilating species within a phase-changing 
carrier fluid is explored by both analytical and numerical means. 
The model takes full account of the dynamic heterogeneity which arises as a result
of the coupling between hydrodynamic transport, dynamic phase-transitions and 
chemical reactions between the participating species, in the presence of a selective 
droplet interface.
Special attention is paid to the dynamic symmetry breaking between the 
mass of the two species entrapped within the expanding droplet as a function of time.
It is found that selective sources are much more effective symmetry breakers than
selective diffusion.
The present study may be of interest for a broad variety of 
advection-diffusion-reaction phenomena with selective fluid interfaces. 
\end{abstract}

\section{Introduction}

The spatial dynamics of mutually annihilating species is a subject of 
wide interdisciplinary concern, with many applications
in chemistry, condensed matter, material science and even cosmology, with the famous 
problem of \textit{baryogenesis}, namely the large asymmetry between matter and antimatter 
observed in the current Universe \cite{kisslinger2016astrophysics,Canetti_2012}.

A pioneering investigation by Toussaint and Wilczek \cite{toussaint1983particle}, pointed out that 
the time asymptotic behaviour of the mutually annihilating species 
($A$ and $B$ for convenience) crucially depends on the initial conditions.
The rationale is quite intuitive: if the species mix, they react and both 
disappear according to the irreversible reaction $A + B \to P$, where 
$P$ denotes a set of product species which contains neither $A$ nor $B$.
If, on the other hand, by some transport mechanism, they manage to demix or segregate
apart, so that the product of their concentrations becomes vanishingly small, then 
annihilation is quenched, thus spawning a chance for both species to survive 
much longer than under homogeneous mixing conditions.

Besides being of great interest on their own right, the details of such survival 
may have plenty of applications in chemistry, material science or biology, an example
 in point being the absorption of drugs within liquid droplets for microfluidics and drug-delivery applications \cite{ weibel2006applications,falcucci2017heterogeneous,montessori2018mesoscale,montessori2019pof}. 

In this paper we consider a specific mechanism of segregation associated with the growth of
droplets within a phase-changing carrier fluid. 
By postulating a selective transport of the two species
across the droplet interface (membrane), we introduce a symmetry-breaking mechanism which
is ultimately responsible for the differential entrapment of the facilitated species 
(the one with higher transmissivity
across the membrane, say $A$) with respect to the inhibited one (say $B$).
The practical question is: how much mass of both species is entrapped in the growing 
and moving droplet as a function of time? 
Once again, this is interesting per-se as a fundamental transport problem in 
dynamically heterogeneous media,
and also for the aforementioned practical purposes.
To the best of our knowledge, no detailed account of the hydrodynamic
complexity associated with a moving and expanding droplet, in the presence of
transport and chemical reaction, has ever been discussed.
This is precisely the aim of the present work, with prospective focus on 
electroweak baryogenesis.

\section{The transport model}

We consider three species $A$, $B$ and $C$, where $C$ is a fluid carrier
undergoing phase-changes, while $A$ and $B$ are passively transported 
by $C$ and mutually annihilate through chemical reactions.

The three species $k=1,2,3$ obey a continuity equation of the form \cite{hundsdorfer2013numerical}

\begin{equation}
\label{eq1}
\partial_t \rho_k + \partial_a  (\rho_k \ u_{k,a}) = r_k + s_k
\end{equation}
where $a=x,y,z$ runs over spatial dimensions and obeys Einstein's summation rule.

In the above $r_k$ and $s_k$ denote the density change rate due to chemical 
reactions and generic sources, respectively.

Species $A$ and $B$ share the same mass, which we set to unity by convention, $m_A=m_B=1$, so that
the number and the mass of the species are the same quantity. 

The two species annihilate at the following rate:
\begin{equation}
r_A = r_B = -\alpha \rho_A \rho_B
\end{equation}
where $\alpha$ is an adjustable reaction parameter.

The $C$ field serves as a carrier for the $A$ and $B$ species and
obeys a non-ideal Navier-Stokes equation:
\begin{equation}
\partial_t (\rho_C \ u_a) + \partial_b P_{C,ab} = s_C
\end{equation}
where
\begin{equation}
P_{C,ab} = \rho_C \ u_a u_b + p_C \delta_{ab} - \sigma_{ab} + \chi \partial_a \rho_C \partial_b \rho_C
\end{equation}
is the non-ideal momentum flux tensor, including the contributions
of inertia, ideal and non-ideal pressure, dissipation and the capillary
forces responsible for the first-order phase transition. 
The term $s_C$ represents an external source of mass.

Species $A$ and $B$ are passively transported by the $C$-field and diffuse across
it with diffusivity coefficients $D_A$ and $D_B$ respectively.

The $A$ and $B$ species experience a selective permeability of the droplet interface, so that an  
excess of $A$ over $B$ accumulates around the interface and further penetrates within the expanding droplet.

The actual amount of mass engulfed within the droplet resulting from such complex transport process
is highly sensitive to the chemical details, as well as to the hydrodynamic evolution of the system. 

Our main aim is to investigate the complex transport phenomena which result from
the dynamic competition between advection, diffusion and reaction processes
taking place in the framework of a phase-changing carrier fluid.

In our stylized model, microscopic symmetry breaking between species $A$ and $B$
is accounted for by two mechanisms: i) different values of the diffusivities and 
ii) different source terms, for species $A$ and $B$, respectively.

\subsection{Selective diffusivity}

The diffusion coefficients are taken in the form:
\begin{equation}
D_k (\rho) = D_k^{in} \theta(\rho-\rho_m) + D_k^{out} \theta(\rho_m-\rho),\;\;\;k=A,B 
\end{equation}
where $\rho_m=(\rho_l+ \rho_v)/2$ is the mean carrier density within (liquid) and
outside (vapour) the droplet, while $\theta(x)$ is the Heavyside step function.
Smoother versions can easily be implemented, but in this work we shall
stay with the discontinuous model.

The $AB$ symmetry-breaking processes are accounted for by choosing a 
different ratio between the inner (within the droplet) and 
outer (outside the droplet) diffusion coefficients of the $A$ and $B$ species, namely, $J_A \ne  J_B$, where we have defined the in-out diffusion jump factors as:
\begin{equation}
J_k \equiv \frac{D_k^{in}}{D_k^{out}},\;k=A,B    
\end{equation}
For convenience, we set the same outer diffusivity for species $A$ and $B$, namely:
\begin{equation}
D_A^{out} = D_B^{out} = D
\end{equation}
In particular, we note that $J<1$, i.e. smaller diffusivity inside the 
droplet than outside, implies a net 
flux towards the interface, due to the diffusive velocity $\vec{u}_D = -\nabla D$. 
This is consistent with the fact that diffusivity is supposed to 
decrease at increasing carrier density.

\subsection{Selective sources}

We shall consider the following source terms
\begin{eqnarray}
s_A(x,y) = s_0 W(\vec{r}-\vec{r}_s)\\
s_B(x,y) = \zeta s_0 W(\vec{r}-\vec{r}_s)
\end{eqnarray} 
where $W(\vec{x})$ is a piece-wise constant centred around the droplet interface, 
and $0 \le \zeta \le 1$ is the symmetry-breaking  parameter.
Such source term models the effect of catalytic reactions acting at 
the droplet surface, although we shall not delve into any detail
of their specific origin.

\section{Numerical set-up}

The transport equations described above are solved by means of a three-species
lattice Boltzmann (LB) scheme \cite{succi2018lattice,BENZI1992145,kruger2017lattice,falcucci2016mapping} (see Appendix). The main reason for using LB is its ability of dealing
with dynamic phase-transitions in a much handier way than solving the Navier-Stokes
equations of non-ideal fluids. 

In the sequel, we introduce the simulation set-up and present 
numerical results for both the selective scenarios described in the 
previous section.

We consider a two-dimensional square with $L=1024$ grid-points per side and run the simulations over a timespan of $10^6$ time-steps, thus covering three decades in space and 
six in time, as it is appropriate for diffusive phenomena.

\subsection{Initial and boundary conditions}

Both species are initialised at the same constant density 
value throughout the computational domain:

\begin{equation}
\rho_A(x,y) = \rho_B(x,y) = \rho_0 = 1
\end{equation}

The initial density of the carrier field is defined as follows:
\begin{eqnarray}
\rho_C(x,y) = \rho_{in} (1 \pm \eta),\; \  &\mbox {in a ball of radius $R_0$}\\
\rho_C(x,y)  = \rho_{out},\;  & \mbox{outside the ball} 
\end{eqnarray}
where $\eta$ is a zero-mean random perturbation with rms $\delta \rho/\rho=0.01$.

We take $\rho_0=1$, $\rho_{in}=0.2$, $\rho_{out}=0.10$ and $R_0=10$.

For the phase transition, we choose $T/T_c=0.04/0.047$, corresponding to a coexistence 
liquid/vapour density ratio of about $0.27/0.027=10$.

The values of $\rho_{out}$ and $\rho_{in}$ determine the duration 
of the growth stage, i.e. the time it takes for the droplet to attain 
the coexistence values of the liquid ($l$) and vapour ($v$) phases.

By mass conservation:
\begin{equation}
\rho_{in} V_{in} + \rho_{out} V_{out} = \rho_l V_l + \rho_v V_v = M 
\end{equation}
where $V_{in}+V_{out} = V_l + V_v = V=L^2$ is the total volume of the system in two dimensions.
 
Clearly, the volume of the droplet grows at increasing the total mass in the system. 
More specifically, the final value of the volume fraction, i.e. the ratio 
of the volume of the liquid droplet to the total volume, is given by:
\begin{equation}
\lambda \equiv \frac{V_l}{V} = \frac{\rho_m-\rho_v}{\rho_l-\rho_v}
\end{equation}
where $\rho_m = M/V$ is the average mass density.

The diameter of the liquid droplet is thus given by:
\begin{equation}
\label{D2L}
D = L \sqrt{ \frac{4}{\pi} \lambda}
\end{equation}
Hence, the maximum droplet diameter, $D=L$, is attained at the a 
volume fraction $\lambda_{max} = \pi/4 \sim 0.78$.

Finally, all species are taken initially at rest, namely:
\begin{equation}
\vec{u}_k(x,y) = 0, \; k=A,B,C
\end{equation}

Full periodicity is assumed across the four boundaries of the simulation box.
 
\subsection{Chemical rates and diffusivities}
 
Collisional time-scales are fixed at $\tau_A = \tau_B = \tau_C=1$, which is the 
fastest timescale in action.
This corresponds to an outer diffusivity and 
carrier viscosity, $D_A=D_B=\nu_C=1/6$ in lattice units (see Appendix).
 
The annihilation rate is taken as $\alpha = 0.1$, corresponding to an 
annihilation timescale $\tau_a = 10$ at unit density $\rho_k=1$.

The diffusion timescale across the membrane is $\tau_{d} = w^2/D$, $w$ being the 
width of the droplet interface. 
Given that in LB simulations $w \sim 5$ lattice units, this 
corresponds to a Damkohler number (diffusive/chemical timescale), $Da \equiv \tau_{d}/\tau_a \sim 15$.
This means, at unit density, annihilation is about $15$ times faster 
than the diffusive time scale, which also means that annihilation 
is effective within the droplet interface.
At densities below $1/15$ the two time scales become comparable, and the
interface becomes chemically transparent.
 

\section{Analytical considerations}

To gain perspective, it is of interest to analyse the homogeneous case, which 
proves amenable to some analytical considerations.
In the homogeneous-symmetric scenario $D_A=D_B=0$, no-phase transitions, no sources 
and symmetric initial conditions, both species decay according 
to the nonlinear homogeneous equation:
\begin{equation}
\frac{d \rho_k}{dt} = -\alpha \rho_k^2
\end{equation}

whose analytical solution reads as follows:
\begin{equation}
\label{1T}
\rho_k(t) = \frac{\rho_{k0}}{1+ \alpha \ \rho_{k0} \ t},\;k=A,B
\end{equation} 

This yields a $\tau_k/t$ decay, where 
\begin{equation}
\tau_k \equiv \frac{1}{\alpha \ \rho_{k0}},\;k=A,B
\end{equation}
is the density-dependent annihilation time-scale.

In the presence of a symmetry-breaking membrane, the two species are expected
to develop different values of $\tau_k$, which we refer as to a {\it dynamic
symmetry breaking}, due to the effect of the selective interface on the species density.
Such dynamic symmetry breaking is expected to occur as soon as the droplet
starts to grow, i.e. it starts to nucleate out of its initial seed of radius $R_0$.
Both species $A$ and $B$ begin to be entrapped within the nucleating droplet 
and their mass within the droplet grows accordingly, as long as the droplet 
growth rate exceeds their annihilation rate. 

As we shall see, such growth is far from monotonic, but characterized instead
by large fluctuations, due to the carrier density waves radiating away from the expanding droplet. 
Such oscillations do not settle down until the droplet condensation has come to an end, i.e
at $t \sim \tau_{con} \equiv R/\dot R$, where $\tau_{con}$ defines the condensation time of the droplet,
i.e. the time it takes for its mass to reach steady-state.

In the long-term, namely at $t \gg \tau_{con}$, the densities of the two species are
expected to settle to constant values inside and outside the droplet, thus leading
to the coexistence of two homogeneous compartments: the droplet and its surrounding environment.  

Since the droplet is homogeneous, the mass of the entrapped species
is expected to follow again a $\tau_k/t$ decay, with two different
values of $\tau_A$ and $\tau_B$, due to the aforementioned dynamic symmetry breaking.

In the sequel, we shall put these qualitative considerations on quantitative grounds
based on the result of extensive numerical simulations. 

\begin{figure}
\centering
\includegraphics[width=0.95\textwidth]{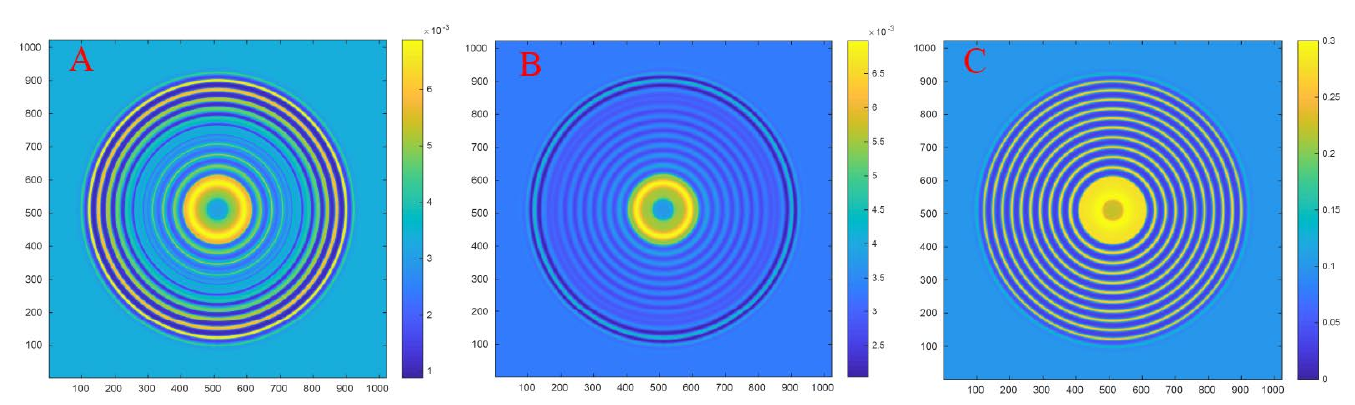}
\caption{\label{fig1} Color plates of the density of the species A, B and C during the growth stage.
Density/pressure waves radiating away from the droplet are clearly visible.
}
\end{figure}

\section{Numerical Results: Selective Diffusion} 

We consider the source-free case $s_A=s_B=0$ and 
define a quantitative symmetry-breaking indicator in the form of the 
diffusivity jump factor,  $J_{AB}=J_B/J_A$. 

We run three representative cases:  $J_{AB}=0.01,0.1,1$, the latter 
denoting the unbroken, symmetric case. 

For the parameters in point, namely $R_0=10$, $\rho_{in}=0.2$, $\rho_{out}=0.1$, the total initial 
mass of the carrier fluid is $M_C(0)= 0.2 \ \pi \ 10^2 + 0.1 \ (1024^2 -\pi \ 10^2) \sim 10^5$, hence the initial
carrier density is $\rho_m \sim 0.1$. 
With $\rho_l = 0.27$ and $\rho_v=0.027$, as dictated by the equation of state, we obtain 
$\lambda \sim 0.61$, which gives a final droplet volume of about $6\cdot 10^5$ lattice units, 
corresponding to a diameter $D \sim 900$ lattice units, pretty close to the maximum 
value that can be attained on a lattice of side $L=1024$ lattice units.

The initial value of the masses is $M_A(0)=M_B(0)= 1024^2 \sim 10^6$, of which
only $100 \pi \sim 314$ lies inside the initial seed droplet. 

For the present parameters, the droplet is found to reach its final size, 
$D \sim 900$ lattice units, after about $\tau_{con} \sim 1.2 \;10^4$ steps, corresponding to 
an average growth rate $\dot R \sim (R_f-R_0)/\tau_{con} \sim (450-10)/1.2 10^4 \sim 0.04$, 
significantly slower than the sound speed, $c_s =1/\sqrt{3}$, both in lattice units.
This implies that the during the growth stage, the expanding droplet 
emanates trains of density waves radiating away from it. 
As we shall see, such density waves are well visible in the simulations.

In Fig.\ref{FIG3} we report the time evolution of the mass $A$, $B$ and $C$
within the droplet, as well as the order (phase-field) parameter:  
\begin{equation}
\phi_{AB} \equiv \frac{M_A-M_B}{M_A+M_B},
\end{equation}
which provides a direct measure of symmetry breaking.
Indeed, by definition, $\phi_{AB}=0$ under symmetric conditions, while   
$\phi_{AB}=\pm 1$ in the full $A$($B$) components, respectively.

As one can see, after a short-term transient, in which both species decrease due to annihilation, 
the $A$ and $B$ masses start to increase, due to the entrapment within the growing droplet.
At about $t \sim 10^3$, large oscillations start to take place, due to the radiation of density(pressure)
waves from the growing droplet, which lead to local condensation and subsequent
evaporation of annular rings around the droplet. 
These rings are well visible in the three snapshots of the carrier density contours, as
reported in the lower insets of panel (a,b), corresponding to three distinct time instants.

It is interesting to notice that in the regime of wild oscillations,
the species $B$ eventually exceeds species $A$, which we tentatively interpret
as a dynamic effect of the presence of the rings. 

Panel b) reports the long-term evolution of the three masses $A$,$B$,$C$, in the 
time frame  $10^3< t<10^6$.
The main result is that the facilitated species $A$ prevails over $B$, but only
by a comparatively small amount.
Indeed, the largest value attained by the order parameter was 
$0.27$ for the case $J_{AB}=0.01$.
Once the droplet settles down, both species start to decay
according to the homogeneous rate $1/t$, although with a slightly different amplitude,
due to the dynamic symmetric breaking which occurred in the condensation stage.

The density contours of species $C$ highlight that the equilibrium spherical shape is reached by the droplet after a very long time-span. 
Notwithstanding the major shape changes, $C$ mass remains constant, and this 
is sufficient for the homogeneous decay $1/t$ to settle down, long before the
droplet attains mechanical equilibrium.

Finite-size effects are also visible, through the reflection of density waves at the boundary.
Indeed, since we work at pretty large values of geometric confinement, $D/L \sim 0.9$, such 
boundary effects are inevitable.

In all the considered cases, symmetry breaking remains comparatively small at all times, 
notwithstanding the large values of the jump coefficients used in the simulations.

\begin{figure}
\centering
\includegraphics[scale=0.95]{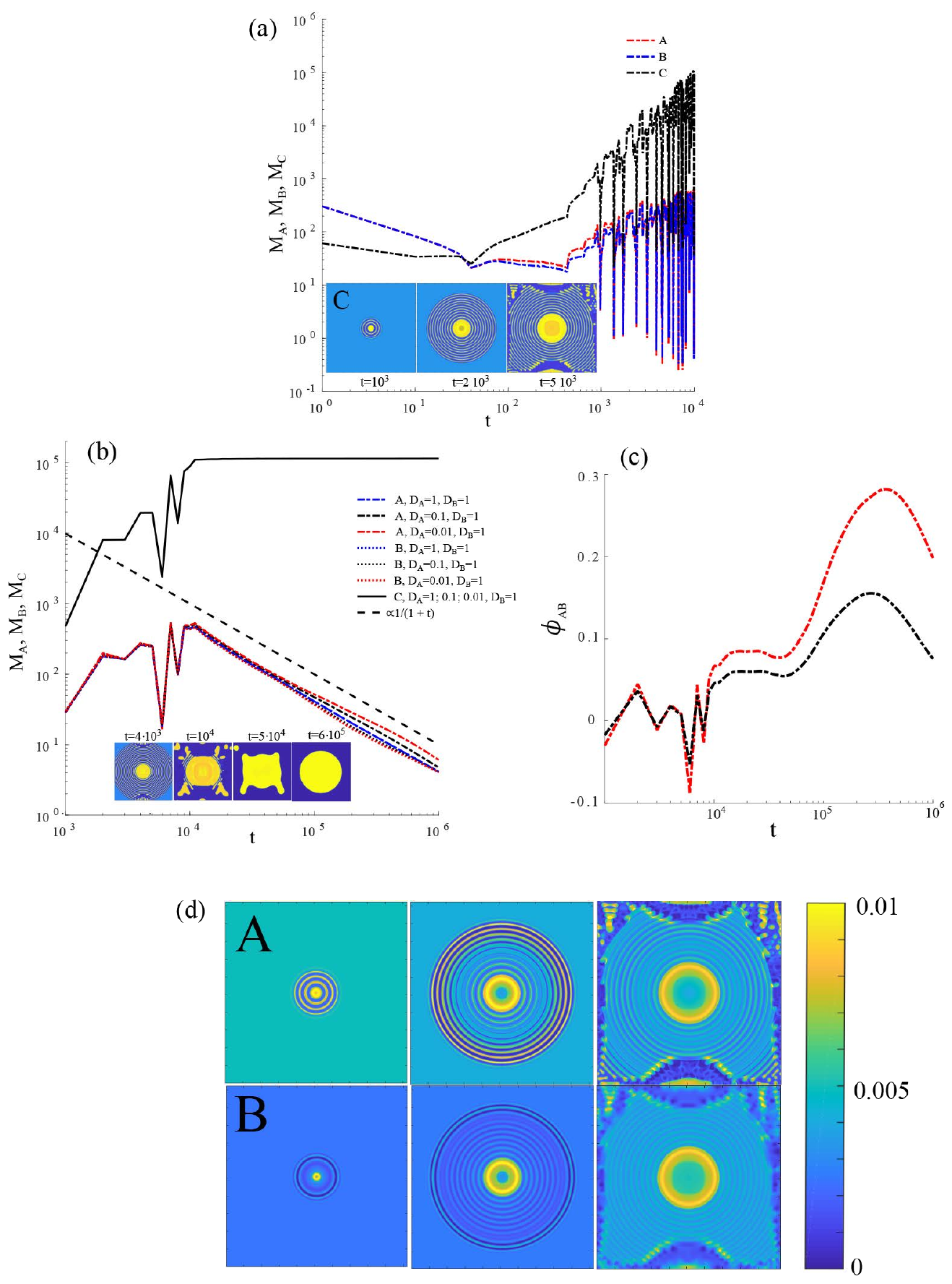}
\caption{
Time evolution of the masses of species $A$, $B$ and $C$ inside the droplet, short 
term (a) and long term (b) for the cases $J_{AB}=0.01$ and $J_{AB}=0.1$ (see legend). The dashed line in panel (b) is a eye-guiding fit $10^7/t$.
In panel (c) we report the long term evolution of the order parameter $\phi_{ab}$ for the cases $J_{AB}=0.01$ (top) and $J_{AB}=0.1$ (bottom). 
The inset in panel (b) reports snapshots of the density contours of the carrier species $C$, while
the density contours of species $A$ and $B$ are reported in panel (d).} 
\end{figure}

\section{Numerical Results: Selective Sources}

In this section, we investigate the effects of selective source terms for the species $A$ and $B$,
by changing the asymmetry source coefficient in the range
$0.4 \le \zeta \le 1$.

The other main parameters are the same as in the previous simulations, except for
$D_A=1,D_B=1$ and $s_0=10^{-3}$.

\begin{figure}
\centering
\includegraphics[scale=0.9,angle=00]{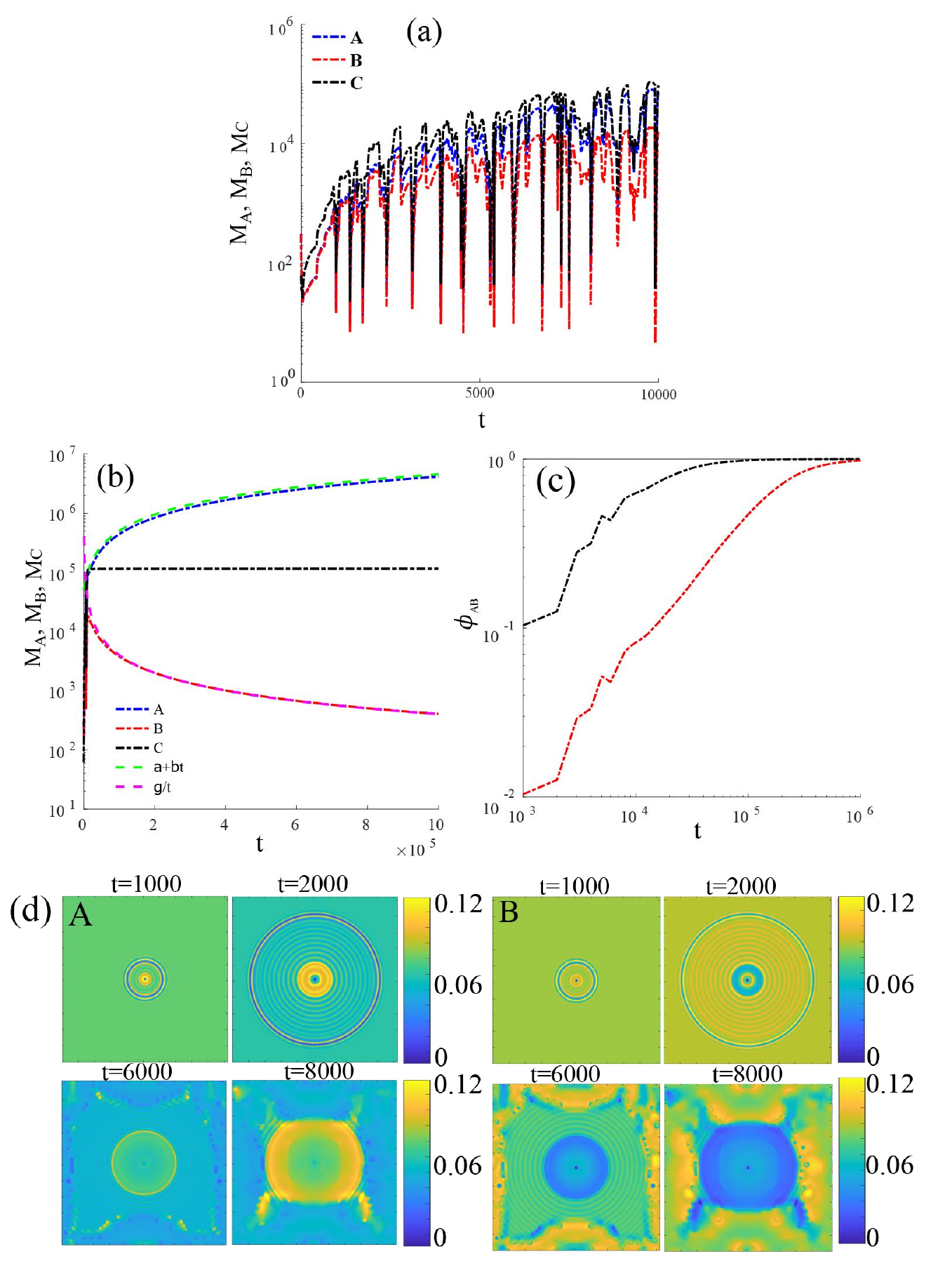}
\caption{ \label{FIG3}
Short (a) and long (b) time evolution of the masses of species $A$, $B$ and $C$ inside the droplet
for the case of selective sources with $\zeta=0.99$.
Panel (c) reports the order parameter $\phi_{AB}$ as a function of time for the 
cases $\zeta=0.99$ (bottom) and $\zeta=0.999$ (top).
Panel (d) reports four snapshots of the density contours of species $A$ (left) and $B$ (right) for the case $\zeta=0.99$.
}
\end{figure}

In Figure \ref{FIG3}, we present the time evolution of the mass 
of species $A$, $B$ and $C$ within the droplet, for the case $\zeta=0.99$.

The short-term behaviour is similar to the case of selective diffusion, although a more substantial symmetric breaking between the $A$ and $B$ masses is observed.

As clearly shown in panel \ref{FIG3}(b), the main difference is the neat separation 
in the long run, due to the fact that any symmetry breaking of the source terms leads 
a secular growth of the facilitated species, $M_A(t) \sim a \ t$, versus 
a homogeneous $b/t$ decay of the unfacilitated one, as we shall discuss shortly.

As a result, the mass ratio goes to zero like $t^{-2}$.

Similarly to the case of selective diffusion, the time asymptotic behaviour sets in long 
before the density configuration in space reaches its mechanical equilibrium, the chief
condition being that the mass of the droplet be stationary in time, regardless of its shape. 

Since the majority species follows a linear trend $M_A(t) \sim t$, while the minority one obeys 
a reciprocal trend $M_B(t) \sim 1/t$, their product remains basically 
asymptotically constant in time, which is indeed confirmed by the numerical results.

As we shall show in the next section, these results can be interpreted in terms of 
analytical solutions of the homogeneous driven case.

\subsection{Homogeneous driven case: analytical model}

The equations of the mass evolution within the droplet for the homogeneous 
(no diffusive fluxes) driven system read as follows:
\begin{eqnarray}
\label{MAB}
\dot M_A = -\alpha R_{AB}  + S_A\\
\dot M_B = -\alpha R_{AB}  + S_B
\end{eqnarray}
where we have defined the global density overlap $R_{AB} = \int \rho_A (x,y) \rho_B(x,y) dx dy$, 
and the global mass inputs per unit time, $S_{k} = \int s_k(x,y) dx dy$, $k=A, B$.

Subtracting the two equations (\ref{MAB}), delivers:
\begin{equation}
\dot M_A - \dot M_B = S_A-S_B,
\end{equation}
which shows that the mass deficit $M_A-M_B$ grows linearly in time.

By multiplying the first by $M_B$, the second by $M_A$ and summing them up, we obtain
\begin{equation}
\frac{d}{dt} (M_A M_B)  = -\alpha R_{AB} (M_A+M_B) + S_B M_A+ S_A M_B 
\end{equation}
Assuming $M_B$ to decay asymptotically to zero, the right hand side is made zero by imposing
\begin{equation}
\label{RAB}
R_{AB} \sim \frac{S_B}{\alpha}
\end{equation}

Next, we write $R_{AB} = \xi_{AB} \frac{M_A M_B}{V_C}$, which defines 
$\xi_{AB}$ as the spatial correlation coefficient, $V_C$ being the droplet volume.

Further expressing the density as a volume average plus a spatial fluctuation,
$\rho_k = M_k/V_C + \tilde \rho_k$, and assuming weak spatial fluctuations,
$\tilde \rho_k V_C/M \ll 1$, the correlation coefficient is made $1$. Thus, the expression (\ref{RAB}) finally yields:

\begin{equation}
\label{MAMB}
M_A M_B = \frac{S_B V_C}{\alpha}  = Const. 
\end{equation}

As anticipated earlier on, this is indeed found to be consistent with the numerical observations.

Summarizing, the time-asymptotic behaviour of the engulfed masses is given by:
\begin{equation}
M_A(t) \sim (S_A-S_B) t
\end{equation}
and
\begin{equation}
M_B(t)  \sim  \frac{S_B V_C}{\alpha M_A(t)}.  
\end{equation}

One can solve explicitly also for the non-asymptotic regime, to obtain:
\begin{eqnarray}
M_A(t) = s_0 V_s \frac{(1-\zeta) t}{2} (\sqrt{1+4 \tau^2/t^2}+1)\\ 
M_B(t) = s_0 V_s \frac{(1-\zeta) t}{2} (\sqrt{1+4 \tau^2/t^2}-1)
\end{eqnarray}
where $V_s=\pi D w$ is the shell volume around the interface.

In the above, we have set 
\begin{equation}
\tau = \tau_s  \; \frac{\zeta^{1/2}}{1-\zeta}
\end{equation}
with $\tau_s \equiv 1/(\alpha s_0)$.
For the current simulations, $\tau_s = 10^4$, fairly close to the 
droplet equilibration time, $\tau_{con}\sim 1.2 \; 10^4$.

The upshot of the above analysis is that the majority species 
grows asymptotically like $(1-\zeta) t$ and the minority species decreases like $\zeta/t$.
As a result, the mass ratio $M_A/M_B$ grows quadratically unbounded in time.

This is potentially far reaching, since it means that even a minuscule asymmetry in the sources 
is destined to give rise to the extinction of the minority species on a times scale 
proportional to $\zeta^{1/2}/(1-\zeta)$, which clearly diverges in the 
symmetric limit $\zeta \to 1$. 
In this limit, $M_A=M_B=\sqrt{\frac{S_BV_C}{\alpha}}$.

A similar treatment goes for the non-homogeneous case, provided the source terms
are augmented with the corresponding diffusive fluxes.

\section{Prospects for electro-weak baryogenesis}

The results presented so far indicate that selective diffusivity is a 
weak symmetry breaker, whereas selective sources are way more effective.

It is therefore of interest to speculate whether the present model can be of  
any use in the context of electro-weak baryogenesis (EWBG) \cite{morrissey2012electroweak}.

To this purpose, let us remind that, so far, we referred to $A$ and $B$ as 
generic mutually annihilating species carried by a phase-changing fluid $C$. \\
Baryogenesis implies the identification
\begin{itemize}
\item $A$ = matter
\item $B$ = antimatter 
\item $C$ = Higgs field
\end{itemize}

Although we refrain from making any claim of quantitative relevance to EWBG, it is nonetheless
of interest to assess the plausibility of present model towards the basic requirements 
laid down by Sakharov, back in the mid sixties \cite{sakharov1991violation}. 

They amount to the following three basic conditions:

\begin{itemize}
\item [] i)   The existence of an explicit baryon-symmetry breaking mechanism, 
\item [] ii)  Violation of C and CP invariance, 
\item [] iii) Thermodynamic non-equilibrium
\end{itemize}

As to i), the baryon symmetry breaking is expressed by the non-unit diffusion jump factor 
$J_{AB}$ across the membrane, or an explicit symmetry breaking 
at the level of source terms, i.e $\zeta \ne 1$.

Item ii) states that the system must be invariant upon a reflection in 
space, say from $x$ to $-x$ across the interface, and charge conjugation. 

Our model is electrically neutral, hence item i) is basically a requirement 
that the density of $A$ at location $x$ be different from the density 
of $B$ at the mirror location $-x$, namely $\rho_A(x) \ne \rho_B(-x)$. This is certainly true once a non-unit diffusivity jump or source asymmetry factor is in action.

Hence the selective models discussed in this work meet both i) and ii) criteria. Finally, thermodynamic non-equilibrium implies that species $A$ and $B$  must depart 
from their local thermodynamic equilibrium, which is certainly true in the presence 
of density gradients across the interface. Thus, even though we do not claim that the model discussed in this work has any 
direct quantitative implications for EWBG, it is nonetheless encouraging to observe 
that it appears to be conceptually compatible with the basic requirements for baryogenesis. To proceed towards a quantitative analysis, several aspects need to be explored in more detail. 
For instance, in the EWBG scenario the Higgs droplet expands much faster 
than the Universe, until it fills it up entirely, whence its alleged pervasiveness 
at the current day \cite{linde1979phase}.

In our model, the droplet stops growing once mass equilibrium is attained, typically
for density ratios around $10$ between the liquid and vapour phases. 
In addition, our computational Universe is static, as opposed to an expanding Universe.
However, both limitations could be significantly mitigated, if needed.

To gain a better understanding of the above issues, it proves useful 
to inspect the physical time and lengthscales of our simulations. 
The time span goes from the onset of EWBG, $t_{EWBG} \sim 10^{-11}$ 
seconds, to the time of the QCD transition, $t_{QCD} \sim 10^{-5}$ seconds.
With one million timesteps, this fixes the lattice timestep to $\Delta t =10$ ps.
The corresponding lattice spacing is $\Delta x = c \Delta t = 3 \; 10^{-3}$ meters, which 
means that we deal with a computational Universe of side $L=3$ meters, and a 
Higgs droplet inflating from about $3$ mm to $3$ meters in diameter.

The droplet growth rate in our simulations is $\dot R \sim 0.04$ in light speed units, 
which is about ten times smaller than the credited wall speed of the true 
Higgs droplets, estimated at $c/2$ \cite{kisslinger2016astrophysics}.

Given that no fine-tuning effort has been spent in customizing the simulations to the 
EBWG scenario, the above figures appear plausible.

Next, let us inspect the values of the matter/antimatter ratio, in our case
the ratio $M_A/M_B$ at the time when the droplet reaches its equilibrium mass. 

\begin{figure}
\centering
\includegraphics[scale=1.2]{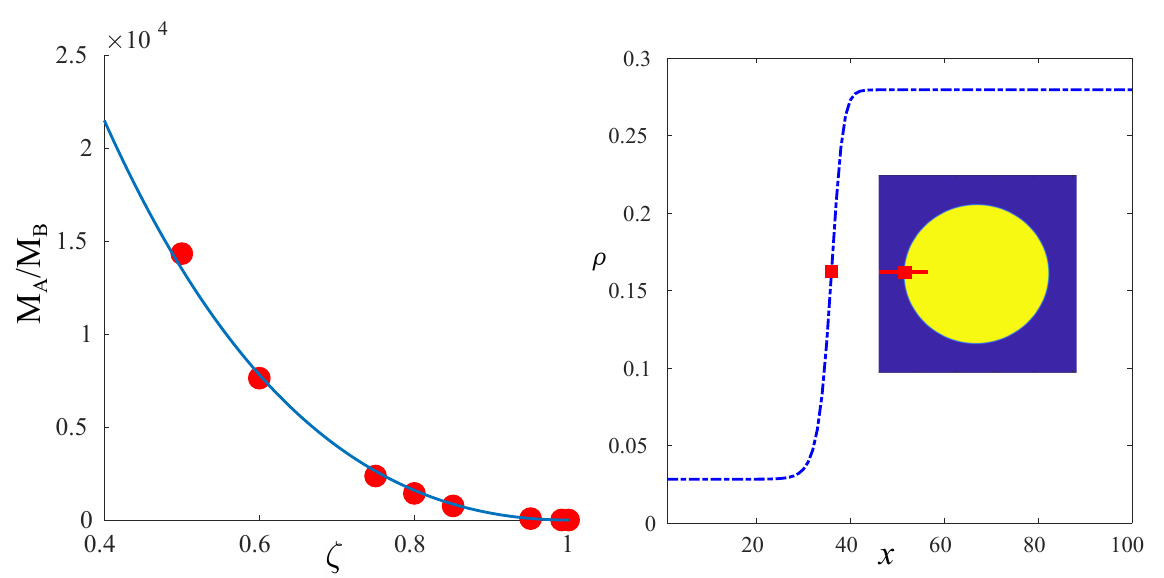}
\caption{\label{FIG5} The ratio $M_A/M_B$  as a function of $\zeta$ at 
the time when the droplet attains its equilibrium mass($t=12000$) (left).
The solid line is the fit $2\cdot \;10^5 e^{-2.3\zeta}\; (1-\zeta)^2$.
On the right side, we show a typical density profile of the carrier species
across the interface, indicated by the horizontal line cutting the 
left edge of the circular droplet.
}
\end{figure}

In Fig. \ref{FIG5} we report the mass ratio $M_A/M_B$ at the end of 
the droplet growth, as a function of the symmetry breaking parameter $\zeta$.

We note that with $\zeta \sim 0.5$, ratios around $10^{4}$ are obtained, which  
extrapolate to $10^{5}$ in the limit $\zeta \to 0$.  
These values are two (one) orders of magnitude above the current value of the 
matter/antimatter ratio in the Universe, which is estimated at about $10^{-6}$. 
Although not visible on the scale of the plot, $\zeta=0.95$ yields 
a mass ratio around $80$, nearly two orders of magnitude, in the 
face of a tiny five percent source asymmetry.

Summarizing, it appears reasonable to speculate that, with proper fine-tuning and extensions,
the present model could prove useful for computational explorations of the semi-classical aspects of strongly non-equilibrium EWBG scenarios. 
The inclusion of quantum effects \cite{riotto1999recent}
may also be feasible through suitable adaptations of the lattice Wigner equation \cite{solorzano2018lattice}.

%
%


\section{Conclusions and outlook}

Summarizing, we have analysed the transport of mutually annihilating species within the flow field
of a passive carrier experiencing a first-order dynamic phase transition.
In particular, we analysed the symmetry-breaking effects on the mass engulfed by the growing droplet
as induced by preferential transport over one species over the other across the droplet interface and
also due to an explicit symmetry breaking of the source terms.

For the source-free case, the evolution proceeds through three dynamic epochs: a 
very short initial $1/t$ decay due to  annihilation, ii) an intermediate stage associated to the droplet
nucleation, in which both masses grow while undergoing large oscillations, with a 
minor prevalence of $A$ over $B$. Finally, a long-term $1/t$ decay in which $A$ 
consistently exceeds $B$ owing to the excess developed in the previous stage.  
This indicates that, although the annihilation rate reaches down to very small 
values,  it is never exactly zero and perfect separation is never achieved. 
As it stands, the model shows that even large diffusivity jump factors across the membrane
do not give rise to any substantial mass asymmetry.
Besides suitable customization of the numerical values, it appears like substantial
mass asymmetry requires additional symmetry-breaking mechanism.

Indeed, the source-driven scenario appears to be much more effective, since 
any nonzero asymmetry between the two 
source terms turns the $1/t$ decay of the majority species into a secular linear growth.
As a result, in the long term, the ratio between minority and majority species decays like $1/t^2$. 
With suitable adaptations, this present model might be able to provide information on the
strongly non-equilibrium spacetime dynamics of the early stage of electroweak baryogenesis. 

\section{Acknowledgements}
The research leading to these results was funded by the European Research Council under the European
Union Horizon 2020 Framework Programme (No. FP/2014-2020)/ERC Grant Agreement No. 739964 (COPMAT).
One of the authors (SS) acknowledges illuminating discussions with 
Gian Francesco Giudice, Gino Isidori, Antonio Riotto and David Spergel.
He also wishes to thank Fabiola Gianotti for arranging a memorable 
visit at CERN, during which part of this work was discussed.

\section{Appendix: The Lattice Boltzmann formulation}

The LB equation takes the following form \cite{succi2018lattice,kruger2017lattice,montessori2018lattice}:
\begin{equation}
f_i^k(\vec{r}+\vec{c}_i \Delta t,t+\Delta t) - f_i^k(\vec{r},t) =
-\frac{\Delta t}{\tau_k} \; (f_i^k-f_i^{k,eq}) + F_i^k \Delta t,\;k=A,B,H
\end{equation}
where $f_i(\vec{r};t)$ represents the probability to find a representative particle of species $k$ 
at the lattice position $\vec{r}$ and time $t$ with the discrete velocity $\vec{c}_i$.
The index $i$ runs over the discrete speeds, $i=0,18$ for the present nineteen-velocity 
three-dimensional lattice.

\begin{figure}
\centering
\includegraphics[scale=0.3]{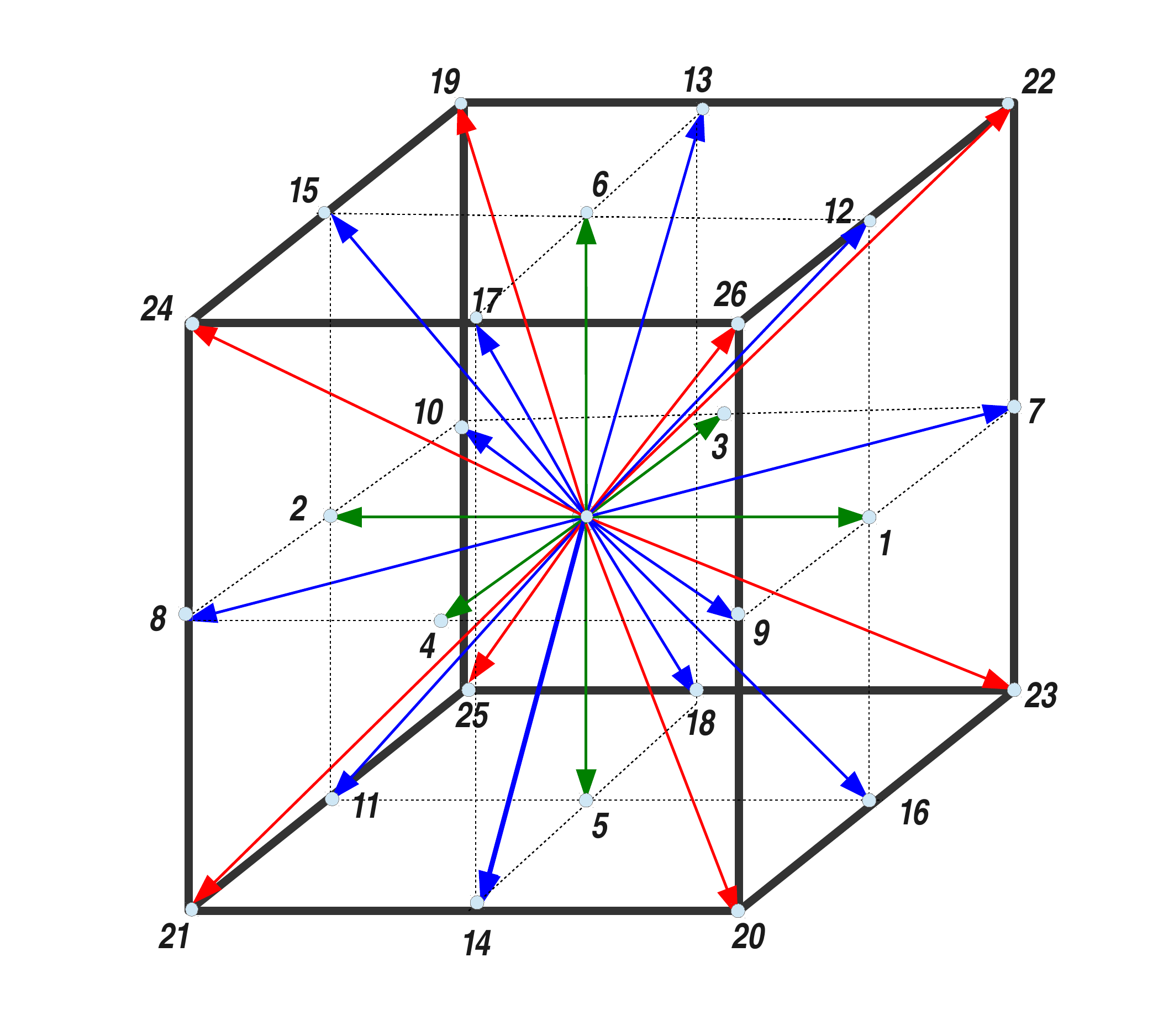}
\caption{The 27 discrete velocity lattice in three spatial dimensions (D3Q27).}
\end{figure}

The local equilibria (a truncated version of Maxwell-Boltzmann distribution)
encode the mass-momentum conservation laws.
At the moment, they are purely classical, but the y can be easily extended to quantum statistics.
In detail, \cite{falcucci2016mapping}: 
\begin{eqnarray}
\label{EQUILS}
f_i^{A,eq} = w_i \rho_A (1+ u_i)\\
f_i^{B,eq} = w_i \rho_B (1+ u_i)\\
f_i^{C,eq} = w_i \rho_C (1+ u_i + q_i/2)\\
\end{eqnarray}
where $u_i = \vec{u} \cdot \vec{c}_i /c_s^2$ and  $q_i = u_i^2 -u^2$, $u$ being the
magnitude of the net flow of the carrier fluid, namely
\begin{equation}
\label{HCUR}
\rho \vec{u} = \sum_i f_i^C \vec{c}_i
\end{equation}
with
\begin{equation}
\rho = \sum_i f_i^C 
\end{equation}
the carrier density.
Finally, $w_i$ is the standard set of weights normalised to unity and 
$c_s^2 = \sum_i w_i c_i^2/d$ is the sound speed in $d$ spatial dimensions. 
In the present lattice $c_s^2=1/3$ (Note that the speed of light is $c=1$ in lattice units).

The transport properties are controlled by the relaxation rate, according to the standard
LB relations, namely:
\begin{eqnarray}
\label{TRA}
D_A = c_s^2   (\tau_A-\Delta t/2),\; D_B = c_s^2   (\tau_B-\Delta t/2), \;\nu_C = c_s^2 (\tau_C-\Delta t/2)
\end{eqnarray}

Note that $A$ and $B$ equilibria conserve only mass, hence they support mass diffusion, whereas carrier $C$ equilibria
conserve momentum as well because the local equilibria contain the self-consistent carrier current, see
Eq. (\ref{HCUR}). Consequently the carrier relaxation rate controls momentum diffusivity, also
known as kinematic viscosity.

For species $A$ and $B$, the forcing terms are set to zero $F_i^A = F_i^B=0$, so that
they obey an ideal equation of state 
$$p_{A,B}=\rho_{A,B} c_s^2.$$
Given that $c_s^2 = 1/3$ in lattice units, where $c=\Delta x/\Delta t=1$ is the light speed. 

The carrier fluid, however, is subject to self-consistent force resulting from potential energy
interactions, according to the standard LB pseudo-potential formulation \cite{shan1993lattice}. 
Consequently, it obeys a non-ideal equation of state of the form (Carnahn-Starling) \cite{carnahan1969equation,montessori2017entropic}
\begin{equation}
\frac{p_C}{\rho_C v_T^2} = \frac{1+ r + r^2 + r^3}{(1-r)^3} - 0.5r
\end{equation}
where $v_T^2=k_B T/m$ and $r$ is the reduced carrier density.
This corresponds to a critical temperature $T_c=0.047$ and density $\rho_c=0.066$ in lattice units.

\subsection{Relaxation time as a function of the carrier density},

In the present paper, we assume a discontinuous jump between the inner and outer 
space, although smoother dependencies could be easily adjusted.  
In the LB scheme, the diffusivity is controlled by the relaxation frequency $\omega \equiv 1/\tau$, hence 
the jump in diffusivity implies a corresponding change of such frequency across the membrane,
Having stipulated $\omega_{out}=1$, the relation (\ref{TRA}) implies: 
$2/\omega_{in} - 1= D_{in}/D_{out} $, that is:
$$
\tau_{out} =1, \;\;\; \tau_{in} = \frac{1}{2} (1 + D_{in}/D_{out}) 
$$
The diffusivity jump $J=D_{in}/D_{out}$ is then the only symmetry-breaking parameter to be varied in the simulations without sources.
For the source-driven case we set, $F_i^k=W_is^k$, see equation (\ref{eq1}) in the text.

\bibliographystyle{plain}

\end{document}